\documentclass[12pt]{amsart}
\usepackage{amsmath, amssymb, amsthm, mathrsfs, mathtools, amscd}
\usepackage[latin1]{inputenc}
\usepackage{hyperref}
\usepackage[all]{xy}

\newtheorem{theorem}{Theorem}[section]
\newtheorem{lemma}[theorem]{Lemma}

\newtheorem{proposition}[theorem]{Proposition}
\newtheorem{definition}[theorem]{Definition}
\newtheorem{remark}[theorem]{Remark}

\newcommand\Sig{\underline{\Sigma}}
\newcommand{\SigM}{\Sig^{\mathcal{M}}}
\newcommand{\SigN}{\Sig^{\mathcal{N}}}

\begin{document}
\title[Abelian subalgebras and Jordan structure]{Abelian subalgebras and the Jordan structure of a von Neumann algebra}

\author{Andreas D\"oring}
\address{Andreas D\"oring, Clarendon Laboratory, Department of Physics, University of Oxford, Parks Road, OX1 3PU, Oxford, UK}
\email{doering@atm.ox.ac.uk}

\author{John Harding}
\address{John Harding, Department of Mathematical Sciences, New Mexico State University, Las Cruces, NM 88003, USA}
\email{jharding@nmsu.edu}

\begin{abstract}
For von Neumann algebras $\mathcal{M},\mathcal{N}$ not isomorphic to $\mathbb{C}\oplus\mathbb{C}$ and without type $I_2$ summands, we show that for an order-isomorphism $f:AbSub~\mathcal{M}\to AbSub~\mathcal{N}$ between the posets of abelian von Neumann subalgebras of $\mathcal{M}$ and $\mathcal{N}$, there is a unique Jordan $*$-isomorphism $g:\mathcal{M}\to \mathcal{N}$ with the image $g[\mathcal{S}]$ equal to $f(\mathcal{S})$ for each abelian von Neumann subalgebra $\mathcal{S}$ of $\mathcal{M}$. The converse also holds. This shows the Jordan structure of a von Neumann algebra not isomorphic to $\mathbb{C}\oplus\mathbb{C}$ and without type $I_2$ summands is determined by the poset of its abelian subalgebras, and has implications in recent approaches to foundational issues in quantum mechanics.
\end{abstract}

\maketitle
%
% 46L10: General theory of von Neumann algebras, 81P05: Quantum theory, general and philosophical. 03G12: Quantum logic. 17C65: Jordan structures on Banach spaces and algebras. 18B25: topoi. 
%


\section {Introduction} 
We consider the question: given a von Neumann algebra $\mathcal{M}$, how much information about $\mathcal{M}$ is encoded in the order structure of its collection of unital abelian von Neumann subalgebras? The set $AbSub~\mathcal{M}$ of such subalgebras, partially ordered by set inclusion, becomes a complete meet semilattice in which every subset that is closed under finite joins has a join. The task is to reconstruct algebraic information about the algebra $\mathcal{M}$ from the order-theoretic structure of $AbSub~\mathcal{M}$. More generally, we are interested in the interplay between these two levels of algebraic structure. 

When $\mathcal{M}$ is abelian, the projection lattice $Proj~\mathcal{M}$ forms a complete Boolean algebra, and one can show that the poset $AbSub~\mathcal{M}$ is isomorphic to the lattice of complete Boolean subalgebras of $Proj~\mathcal{M}$. Modifying a result of Sachs \cite{Sachs} that every Boolean algebra is determined by its lattice of all subalgebras, to show each complete Boolean algebra is determined by its lattice of complete subalgebras, one can then obtain that $Proj~\mathcal{M}$ is determined by $AbSub~\mathcal{M}$. That $\mathcal{M}$ is determined by $Proj~\mathcal{M}$ is a consequence of the spectral theorem. 

For a {\em non-abelian} von Neumann algebra, the situation is more complicated. \mbox{Reconstruction} of the non-commutative product in $\mathcal{M}$ will not generally be possible as there are non-isomorphic von Neumann algebras having the same Jordan product, hence exactly the same posets of unital abelian subalgebras. However, we will show that the order structure of $AbSub~\mathcal{M}$ does determine $\mathcal{M}$ as a Jordan algebra up to (Jordan) $*$-isomorphism. This means that the poset $AbSub~\mathcal{M}$ encodes a substantial amount of algebraic information about $\mathcal{M}$. The proof goes along the same lines as the abelian case, using a result of \cite{HardingNavara} that an orthomodular lattice is determined by its poset of Boolean subalgebras. In fact, our result is somewhat stronger than we described. 
\vspace{2ex}

\noindent {\bf Theorem } {\em Suppose $\mathcal{M}, \mathcal{N}$ are von Neumann algebras without type $I_2$ summands and $f:AbSub~\mathcal{M}\to AbSub~\mathcal{N}$ is an order-isomorphism. Then there is a unique Jordan $*$-isomorphism $F:\mathcal{M}\to \mathcal{N}$ with $f(\mathcal{S})$ equal to the image $F[\mathcal{S}]$ for each $\mathcal{S}$.}
\vspace{2ex}

This result is particularly interesting with respect to the so-called topos \mbox{approach} to the formulation of physical theories \cite{DI(1),DI(2),DI(3),DI(4),DI08,HLS09}, where a mathematical reformulation of algebraic quantum theory is suggested. \mbox{For a von Neumann algebra} $\mathcal{M}$, one considers the poset $AbSub~\mathcal{M}$ of its abelian subalgebras and the topos of presheaves over this poset. The idea is that each abelian subalgebra represents a `classical perspective' on the quantum system. By taking all classical perspectives together, one obtains a complete picture of the quantum system. Mathematically, this corresponds to considering the poset $AbSub~\mathcal{M}$ and presheaves over it. These presheaves form the topos associated with the quantum system. The so-called spectral presheaf $\SigM$, whose components are the Gelfand spectra of the abelian von Neumann subalgebras of $\mathcal{M}$, plays a key role in the topos approach. Physically, the spectral presheaf is interpreted as a generalized state space for the quantum system described by the algebra $\mathcal{M}$. Mathematically, $\SigM$ is a kind of spectrum of the non-abelian von Neumann algebra $\mathcal{M}$. It becomes clear that, from the perspective of the topos approach, it is very relevant to see how much information about the algebra $\mathcal{M}$ can be extracted from the poset $AbSub~\mathcal{M}$.

Since the appearance of the draft of this manuscript on ArXiv \cite{DHArXiv}, several related manuscripts and papers have arisen. In \cite{HamhalterC*} a related task is undertaken for the poset of abelian subalgebras of a $C^*$-algebra, and in \cite{HamhalterTurilova} the matter is considered from the viewpoint of associative subalgebras of a Jordan algebra. In \cite{Doe} applications to the topos approach to physical theories are considered further. In particular, it is shown that if $M,N$ are von Neumann algebras with no direct summands of type $I_2$, then there is a Jordan $*$-isomorphism $F:\mathcal{M}\to \mathcal{N}$ if and only there is an isomorphism $\Phi:\SigN\to\SigM$ between their spectral presheaves in the opposite direction.

\section{Preliminaries}

For a complex Hilbert space $H$, let $\mathcal{B}(H)$ be the $C^*$-algebra of all bounded operators on $H$. For a subset $\mathcal{S}\subseteq \mathcal{B}(H)$, the commutant $\mathcal{S}'$ is the set of all elements of $\mathcal{B}(H)$ that commute with each member of $\mathcal{S}$. A von Neumann algebra is a subset $\mathcal{M}\subseteq \mathcal{B}(H)$ with $\mathcal{M}=\mathcal{M}''$. For a von Neumann algebra $\mathcal{M}$, we use $Proj~\mathcal{M}$ for the set of projections in $\mathcal{M}$. The following well-known result \cite[pg.~69]{Kalmbach} will be used repeatedly. 

\begin{proposition}
For $\mathcal{M}$ a von Neumann algebra, $\mathcal{M}=(Proj~\mathcal{M})''$.
\end{proposition}

For any von Neumann algebra $\mathcal{M}$ the projections $Proj~\mathcal{M}$ form a complete orthomodular lattice (abbreviated: \textsc{oml}). Our primary interest lies in subalgebras of von Neumann algebras, subalgebras of their projection lattices, and relationships between these and the original von Neumann algebra. We require several definitions. 

\begin{definition}
A von Neumann subalgebra of a von Neumann algebra $\mathcal{M}$ is a subset $\mathcal{S}\subseteq \mathcal{M}$ that is itself a von Neumann algebra. 
\end{definition} 

We will only consider von Neumann subalgebras $\mathcal{S}\subseteq\mathcal{M}$ such that the unit elements in $\mathcal{S}$ and $\mathcal{N}$ coincide. (In particular, we will not consider subalgebras of the form $\hat{P}\mathcal{M}\hat{P}$ for a non-trivial projection $\hat P\in\mathcal{M}$.)
%Note that the unit $1$ of $\mathcal{B}(H)$ belongs to the commutator of any set, hence to each von Neumann algebra. So a von Neumann subalgebra $\mathcal{S}$ of $\mathcal{M}$ has the same unit as $\mathcal{M}$, or in other words is a unital subalgebra.
We remark that being a von Neumann subalgebra is equivalent to being a unital $C^*$-subalgebra that is closed in the $\sigma$-weak topology, equivalent to being a unital $C^*$-subalgebra that is closed under monotone joins \cite[pg.~101--110]{AS}. 

\begin{definition}
For a von Neumann algebra $\mathcal{M}$, we let $Sub~\mathcal{M}$ be the set of all von Neumann subalgebras of $\mathcal{M}$ ordered by set inclusion; $AbSub~\mathcal{M}$ be the set of abelian von Neumann subalgebras of $\mathcal{M}$ ordered by set inclusion; and $FAbSub~\mathcal{M}$ be the set of all abelian subalgebras of $\mathcal{M}$ that contain only finitely many projections, ordered by set inclusion. 
\end{definition}

We note that $Sub~\mathcal{M}$ is a complete lattice, with meets given by intersections. The join of a family $(\mathcal{S}_i)_{i\in I}$ of subalgebras is the weak closure of the algebra generated by the algebras $\mathcal{S}_i$, $i\in I$. Analogously, $AbSub~\mathcal{M}$ is a complete meet semilattice where every subset that is closed under finite joins has a join, and $FAbSub~\mathcal{M}$ is a complete meet semilattice where every meet is essentially finite. Yet, neither $AbSub~\mathcal{M}$ nor $FAbSub~\mathcal{M}$ have a top element if $\mathcal{M}$ is non-abelian, so empty meets do not exist in these posets.

\begin{definition}
For an \textsc{oml} $L$, we let $Sub~L$ be the set of all subalgebras of $L$; $BSub~L$ be the set of Boolean subalgebras of $L$, and $FBSub~L$ be the set of finite Boolean subalgebras of $L$, all partially ordered by set inclusion. If $L$ is complete we let $CSub~L$ be the set of complete subalgebras of $L$, meaning subalgebras that are closed under arbitrary joins and meets from $L$, and $CBSub~L$ be the set of complete Boolean subalgebras of $L$. Again, these are considered as posets, partially ordered by set inclusion.
\end{definition} 

For a von Neumann algebra $\mathcal{M}$ we can use the associative, but not necessarily commutative, product on $\mathcal{M}$ to define a commutative, but not necessarily associative product $\circ$ on $\mathcal{M}$, called the Jordan product, by setting 
\[a\circ b = \frac{1}{2}(ab + ba).\]

Suppose $\varphi$ is a map between von Neumann algebras that is linear, bijective, and preserves the involution (adjoint) $*$. We say $\varphi$ is a $*$-isomorphism if it satisfies $\varphi(ab)=\varphi(a)\varphi(b)$; a $*$-antiisomorphism if it satisfies $\varphi(a b)=\varphi(b)\varphi(a)$; and a Jordan isomorphism if it satisfies $\varphi(a\circ b) = \varphi(a)\circ\varphi(b)$. The following is well known \cite{KRII,Takesaki}.

\begin{proposition}
Every Jordan isomorphism $\eta:\mathcal{M}\rightarrow\mathcal{N}$ between von Neumann algebras $\mathcal{M},\mathcal{N}$ can be decomposed as the sum of a $*$-isomorphism and a $*$-anti-isomorphism. 
\end{proposition}

More concretely, there are central projections $\hat{P}_1,\hat{P}_2\in\mathcal{M}$ and $\hat{Q}_1,\hat{Q}_2\in\mathcal{N}$ such that $\mathcal{M}$ and $\mathcal{N}$ are unitarily equivalent to $\mathcal{M}\hat{P}_1\oplus\mathcal{M}\hat{P}_2$ and $\mathcal{N}\hat{Q}_1\oplus\mathcal{N}\hat{Q}_2$, respectively, and $\eta|_{\mathcal{M}\hat{P}_1}:\mathcal{M}\hat{P}_1\rightarrow\mathcal{N}\hat{Q}_1$ is a $*$-isomorphism, while $\eta|_{\mathcal{M}\hat{P}_2}:\mathcal{M}\hat{P}_2\rightarrow\mathcal{N}\hat{Q}_2$ is a $*$-antiisomorphism.

It follows from \cite{Connes} that there is a von Neumann algebra that is not $*$-isomorphic to its opposite, hence these two von Neumann algebras are Jordan isomorphic, but not $*$-isomorphic. So there can be two different associative noncommutative products on a weakly closed set of operators, giving different von Neumann algebras, but the same Jordan structure. So the associative noncommutative product on a von Neumann algebra cannot be recovered from the lattice of its subalgebras as a von Neumann algebra and its opposite will have precisely the same subalgebras. However, we will see that in the absence of type $I_2$ summands (and excluding the case $\mathcal{M}=\mathbb{C}\oplus\mathbb{C}$), the Jordan structure can be recovered. The following result by Dye \cite{Dye}, see also \cite[Theorem~8.1.1]{Hamhalter}, will be of key importance. We note that the uniqueness in the version of this result given below follows from the spectral theorem. 

\begin{theorem}
Suppose $\mathcal{M},\mathcal{N}$ are von Neumann algebras without type $I_2$ summands. Then for any \textsc{oml}-isomorphism $\psi:Proj~\mathcal{M}\to Proj~\mathcal{N}$ there is a unique Jordan $*$-isomorphism $\Psi:\mathcal{M} \to \mathcal{N}$ with $\Psi(p)=\psi(p)$ for each projection $p$ of $\mathcal{M}$. 
\label{theorem:Jordan}
\end{theorem}

The reader should consult \cite{AS,Fillmore,KRI,KRII,Takesaki} for basics on von Neumann algebras, \cite{Birkhoff} for lattice theory, and \cite{Kalmbach} for \textsc{oml}s. 

\section{Main result}

\begin{lemma} 
Let $\mathcal{M}$ be  a von Neumann algebra. Then there is an order-isomorphism $\Psi:FAbSub~\mathcal{M} \to FBSub~(Proj~\mathcal{M})$ defined by setting $\Psi \mathcal{S} = \mathcal{S}\cap Proj~\mathcal{M}$. 
\label{lem:bob}
\end{lemma}

\begin{proof} It follows from \cite[Theorem~2.104]{AS} that the projections of any abelian subalgebra of $\mathcal{M}$ form a Boolean subalgebra of $Proj~\mathcal{M}$. So $\Psi$ is indeed a map from $FAbSub~\mathcal{M}$ to $FBSub~(Proj~\mathcal{M})$. Clearly $\Psi$ is order-preserving. Suppose $\Psi \mathcal{S} \subseteq \Psi \mathcal{T}$. As $\mathcal{S}$ is a von Neumann algebra $\mathcal{S}=(Proj~\mathcal{S})''$, and similarly for $\mathcal{T}$. Therefore $\mathcal{S} = (\Psi \mathcal{S})'' \subseteq (\Psi \mathcal{T})'' = \mathcal{T}$, showing $\Psi$ is an order-embedding. 

Suppose $B$ is a finite Boolean algebra of projections in $\mathcal{M}$ with atoms $p_1,\ldots,p_n$, and consider the map $\Lambda : \mathbb{C}^n\to \mathcal{M}$ defined by setting $\Lambda(\lambda_1,\ldots,\lambda_n) = \sum_1^n\lambda_i p_i$. One easily sees $\Lambda$ is a normal, unital $*$-isomorphism, so by \cite[Lemma~2.100]{AS} its image $\mathcal{S}$ is a von Neumann subalgebra of $\mathcal{M}$. Clearly $\mathcal{S}$ is an abelian, has finitely many projections, and $\Psi \mathcal{S} = B$. So $\Psi$ is onto. 
\end{proof}

\begin{remark}{\em 
While not needed for our results, it is natural to consider several questions related to the above result. It is easy to see that as above there is an order-embedding $\Psi:Sub~\mathcal{M}\to CSub~(Proj~\mathcal{M})$ that preserves all meets. A simple example with $\mathcal{M}$ being the bounded operators on $\mathbb{C}^2$ shows this map need not preserve joins or be onto. A more difficult argument, using the notion of Bade subalgebras and results from \cite{Ricker}, shows there is an order-isomorphism $\Psi:AbSub~\mathcal{M}\to CBSub~(Proj~\mathcal{M})$. The result above follows from this more general one, but is not needed here.  
}
\end{remark}

\begin{lemma}
For \textsc{oml}s $L$, $M$, each order-isomorphism $\mu:FBSub~L\to FBSub~M$ extends uniquely to an isomorphism $\bar{\mu}:BSub~L\to BSub~M$.  
\label{lem:fred}
\end{lemma}

\noindent {\bf Proof. } We define an ideal of $FBSub~L$ to be a downset $I$ of $FBSub~L$ where any two elements of $I$ have a join, and this join belongs to $I$. For any element $x$ of $BSub~L$, we have $x\!\downarrow \cap FBSub~L = \{z\in FBSub~L:z\subseteq x\}$ is an ideal of $FBSub~L$ and the join of this ideal in $BSub~L$ is equal to $x$. Further, each ideal of $FBSub~L$ is of this form as can be easily seen from the compactness of finitely generated subalgebras in a subalgebra lattice. 

Define $\bar{\mu}$ by setting $\bar{\mu}(x) = \bigvee \mu[ x\!\downarrow\cap FBSub~L]$. This join is well defined as the image under the isomorphism $\mu$ of an ideal is an ideal. Clearly $\bar{\mu}$ is order preserving. Suppose $\bar{\mu}(x)\leq \bar{\mu}(y)$. Then for each $z\in x\!\downarrow\cap FBSub~L$ we have $\mu(z)\leq \bigvee \mu[y\!\downarrow\cap FBSub~L]$. Compactness then yields $z\leq y$ for each such $z$, giving $x\leq y$. Thus $\bar{\mu}$ is an order-embedding. To see $\bar{\mu}$ is onto, note each element $w$ of $BSub~M$ is the join of an ideal $J$ of $FBSub~M$. The preimage $\mu^{-1}[J]$ is an ideal of $FBSub~L$, so has a join $x$ in $BSub~L$. Then $\bar{\mu}(x)=w$, showing $\bar{\mu}$ is onto. 

Clearly $\bar{\mu}$ extends $\mu$. If $\tilde{\mu}$ is another isomorphism from $BSub~L$ to $BSub~M$ extending $\mu$, then $\tilde{\mu}$ preserves joins, so $\tilde{\mu}(x) = \bigvee \mu[x\!\downarrow\cap FBSub~L] = \bar{\mu}(x)$. $\Box$
\vspace{2ex}

We are ready to provide our main result. 

\begin{theorem}
Suppose $\mathcal{M}, \mathcal{N}$ are von Neumann algebras not isomorphic to $\mathbb{C}\oplus\mathbb{C}$ and without type $I_2$ summands, and suppose that $f:AbSub~\mathcal{M}\to AbSub~\mathcal{N}$ is an order-isomorphism. Then there is a unique Jordan $*$-isomorphism $F:\mathcal{M}\to \mathcal{N}$ with $f(\mathcal{S})$ equal to the image $F[\mathcal{S}]$ for each $\mathcal{S}$.
\end{theorem}

\noindent {\bf Proof. } Consider a series of mappings, starting with the given 
\[AbSub~\mathcal{M} \stackrel{f}{-\!\!\!-\!\!\!\longrightarrow} AbSub~\mathcal{N}.\]
We then restrict this to $FAbSub~\mathcal{M}$. Note that the members of $FAbSub~\mathcal{M}$ are precisely those members of $AbSub~\mathcal{M}$ that have only finitely many elements beneath them, and similarly for $FAbSub~\mathcal{N}$. Thus this restriction $g$ is also an order-isomorphism. 
\[FAbSub~\mathcal{M} \stackrel{g}{-\!\!\!-\!\!\!\longrightarrow} FAbSub~\mathcal{N}.\]

Lemma~\ref{lem:bob} gives order-isomorphisms $\Psi_{\mathcal{M}}:FAbSub~\mathcal{M}\to$ \linebreak[4] $FBSub~(Proj~\mathcal{M})$ and $\Psi_{\mathcal{N}}:FAbSub~\mathcal{N}\to FBSub~(Proj~\mathcal{N})$ given by $\Psi_{\mathcal{M}}(\mathcal{S}) = \mathcal{S}\cap Proj~\mathcal{M}$ and $\Psi_{\mathcal{N}}(\mathcal{T}) = \mathcal{T}\cap Proj~\mathcal{N}$. It follows there is a unique order-isomorphism $h$ as below with $h(\mathcal{S}\cap Proj~\mathcal{M}) = g(\mathcal{S})\cap Proj~\mathcal{N}$ for each $\mathcal{S}\in FAbSub~\mathcal{M}$. 
\[FBSub~(Proj~\mathcal{M}) \stackrel{h}{-\!\!\!-\!\!\!\longrightarrow} FBSub~(Proj~\mathcal{N}).\]
Then by Lemma~\ref{lem:fred} this extends uniquely to an order-isomorphism 
\[BSub~(Proj~\mathcal{M}) \stackrel{j}{-\!\!\!-\!\!\!\longrightarrow} BSub~(Proj~\mathcal{N}).\]

The main result of \cite{HardingNavara} says that if $L,M$ are \textsc{oml}s without any 4-element blocks (a block is a maximal Boolean subalgebra), then for any order-isomorphism $\alpha:BSub~L\to BSub~M$ there is a unique \textsc{oml}-isomorphism $\beta:L\to M$ with $\alpha(D) = \beta[D]$ for each Boolean subalgebra $D$ of $L$. As $\mathcal{M},\mathcal{N}$ are neither isomorphic to $\mathbb{C}\oplus\mathbb{C}$ nor to $\mathcal{B}(\mathbb{C}\oplus\mathbb{C})$ (the latter is a von Neumann algebra of type $I_2$), there are no 4-element blocks in $Proj~\mathcal{M}$ or $Proj~\mathcal{N}$. So the map $j$ defined above gives a unique map $k$ as shown below with $j(D) = k[D]$ for each Boolean subalgebra $D$ of $Proj~\mathcal{M}$. 
\[Proj~\mathcal{M} \stackrel{k}{-\!\!\!-\!\!\!\longrightarrow} Proj~\mathcal{N}.\]
Finally, Theorem~\ref{theorem:Jordan} gives a unique Jordan $*$-isomorphism $F$ as below extending $k$. 
\[\mathcal{M} \stackrel{F}{-\!\!\!-\!\!\!\longrightarrow} \mathcal{N}.\]
\vspace{-0.5ex}

\noindent {\bf Claim 1 : } If $\mathcal{S}\in FAbSub~\mathcal{M}$ then $f(\mathcal{S})\cap Proj~\mathcal{N} = F[\mathcal{S}]\cap Proj~\mathcal{N}$. 
\vspace{2ex}

\noindent {\bf Proof : } To see this, note that for such $\mathcal{S}$, 
\begin{eqnarray*}
f(\mathcal{S})\cap Proj~\mathcal{N} &=& g(\mathcal{S})\cap Proj~\mathcal{N}\\
&=& h(\mathcal{S}\cap Proj~\mathcal{M})\\
&=& j(\mathcal{S}\cap Proj~\mathcal{M})\\
&=& k[\mathcal{S}\cap Proj~\mathcal{M}]\\
&=& F[\mathcal{S}\cap Proj~\mathcal{M}]\\
&=& F[\mathcal{S}]\cap Proj~\mathcal{N}
\end{eqnarray*}
The first equality follows as $g$ is the restriction of $f$; the second by the definition of $h$; the third as $j$ extends $h$; the fourth by the definition of $k$; the fifth as $F$ extends $k$; and the sixth as $F$ restricts to a bijection between $Proj~\mathcal{M}$ and $Proj~\mathcal{N}$. $\Box$
\vspace{2ex}

\noindent {\bf Claim 2 : } If $\mathcal{S}\in AbSub~\mathcal{M}$, then $F[\mathcal{S}]\in AbSub~\mathcal{N}$. 
\vspace{2ex}

\noindent {\bf Proof : } As $F$ is Jordan and $\mathcal{S}$ is abelian, by \cite[pg.~187]{Takesaki} the restriction $F|\mathcal{S}$ preserves the associative product. By \cite[pg.~189]{AS} $F$ is a unital order-isomorphism, so it preserves monotone joins, and as $\mathcal{S}$ is a von Neumann subalgebra of $\mathcal{M}$, the identical embedding of $\mathcal{S}$ into $\mathcal{M}$ preserves monotone joins. So the composite $F|\mathcal{S}$ preserves monotone joins, hence is a normal unital one-one $*$-homomorphism of $\mathcal{S}$ into $\mathcal{N}$. So by \cite[Lemma~2.100]{AS} the image $F[\mathcal{S}]$ is a von Neumann subalgebra of $\mathcal{N}$ that is clearly abelian. $\Box$
\vspace{2ex}

\noindent {\bf Claim 3 : } If $\mathcal{S}\in AbSub~\mathcal{M}$, then $f(\mathcal{S})=F[\mathcal{S}]$. 
\vspace{2ex}

\noindent {\bf Proof : } A projection $p$ belongs to $F[\mathcal{S}]$ if, and only if, it belongs to $F[\mathcal{U}]$ for some $\mathcal{U}\subseteq \mathcal{S}$ with $\mathcal{U}\in FAbSub~\mathcal{M}$. The proof is essentially that of Lemma~\ref{lem:bob}. By Claim~1, this is equivalent to $p$ belonging to $f(\mathcal{U})$ for some $\mathcal{U}\subseteq \mathcal{S}$ with $\mathcal{U}\in FAbSub~\mathcal{M}$. As the members of $FAbSub~\mathcal{M}$ are exactly the members of $AbSub~\mathcal{M}$ with finitely many elements beneath them, it follows from $f$ being an order-isomorphism that $\mathcal{T}=F[\mathcal{U}]$ for some $\mathcal{U}\subseteq \mathcal{S}$ with $\mathcal{U}\in FAbSub~\mathcal{M}$ if, and only if, $\mathcal{T}\subseteq f(\mathcal{S})$ and $\mathcal{T}\in FAbSub~\mathcal{N}$. So $p$ belonging to $F[\mathcal{S}]$ is equivalent to $p$ belonging to $\mathcal{T}$ for some $\mathcal{T}\subseteq f(\mathcal{S})$ with $\mathcal{T}\in FAbSub~\mathcal{N}$, so equivalent to $p$ belonging to $f(\mathcal{S})$. By Claim~2, $f(\mathcal{S})$ and $F[\mathcal{S}]$ are von Neumann subalgebras of $\mathcal{N}$, and they contain the same projections, so $f(\mathcal{S})=F[\mathcal{S}]$. $\Box$. 
\vspace{2ex}

To conclude the proof of the theorem, it remains to show uniqueness. Suppose $G:\mathcal{M}\to \mathcal{N}$ is a Jordan $*$-isomorphism with $f(\mathcal{S})=G[\mathcal{S}]$ for each $\mathcal{S}\in AbSub~\mathcal{M}$. Using the spectral theorem, it follows that two Jordan $*$-isomorphisms from $\mathcal{M}$ to $\mathcal{N}$ agreeing on the projections must be equal. So it is enough to show that $F$ and $G$ agree on $Proj~\mathcal{M}$. From the uniqueness of the result in \cite{HardingNavara} it is enough to show $F[D] = G[D]$ for each Boolean subalgebra $D$ of $Proj~\mathcal{M}$, and by the uniqueness in Lemma~\ref{lem:fred} it is enough to show this for finite Boolean subalgebras $D$ of $Proj~\mathcal{M}$. Using Lemma~\ref{lem:bob}, it is then enough to show $F[\mathcal{S}\cap Proj~\mathcal{M}] = G[\mathcal{S}\cap Proj~\mathcal{M}]$ for each $\mathcal{S}\in FAbSub~\mathcal{M}$, and this is a direct consequence of the assumption that $F[\mathcal{S}] = G[\mathcal{S}]$. This shows $F=G$, and concludes the proof of the theorem. $\Box$
\vspace{2ex}

We finally observe that the converse of the above result also holds (in fact, for arbitrary von Neumann algebras):

\begin{proposition}
Let $\mathcal{M},\mathcal{N}$ be von Neumann algebras, and let $F:\mathcal{M}\rightarrow\mathcal{N}$ be a Jordan $*$-isomorphism. Then $F$ induces a unique order isomorphism $f:AbSub~\mathcal{M}\rightarrow AbSub~\mathcal{N}$ with $f(S)$ equal to the image $F[\mathcal{S}]$ for each $\mathcal{S}$.
\end{proposition}

\noindent {\bf Proof :} It is well-known that a Jordan $*$-homomorphism $F:\mathcal{M}\rightarrow\mathcal{N}$ between von Neumann algebras preserves commutativity (see e.g. \cite{Takesaki,KRI2}), so $F$ maps abelian subalgebras of $\mathcal{M}$ to abelian subalgebras of $\mathcal{N}$ in a bijective and order-preserving way. Hence, we obtain an order-isomorphism $f:AbSub~\mathcal{M}\rightarrow AbSub~\mathcal{N}$. $\Box$

%It is well-known that a Jordan homomorphism $\Phi:R_1\rightarrow R_2$ between rings preserves commutativity provided that the associative ring generated by $\Phi(R_1)$ has no non-zero nilpotent elements in its center (\cite{JR50}, Corollary 1). Since $F:\mathcal{M}\rightarrow\mathcal{N}$ is a Jordan $*$-isomorphism by assumption, the ring generated by $F(\mathcal{M})$ is $\mathcal{N}$, which has no nilpotent elements in its center, so $F$ preserves commutativity and maps abelian subalgebras of $\mathcal{M}$ to abelian subalgebras of $\mathcal{N}$ in a bijective and order-preserving way. Hence, we obtain an order-isomorphism $f:AbSub~\mathcal{M}\rightarrow AbSub~\mathcal{N}$. $\Box$

\section{Conclusions}

There remain several directions for further research. First, it would be of interest to see if the Jordan structure of a $C^*$-algebra is determined by its poset of abelian $C^*$-subalgebras. In this direction we remark that it is known that the lattice of $C^*$-subalgebras of an abelian $C^*$-algebra determines the $C^*$-algebra \cite[Theorem~11]{Mendivil}. Perhaps \cite{Rosicky} may also be related to this question. \cite{vdBHeu10} is concerned with abelian subalgebras of partial $C^*$-algebras and von Neumann algebras.

For a different direction, one might consider the matter of adding additional information to the poset $AbSub~\mathcal{M}$ in hopes of recovering the full von Neumann structure of $\mathcal{M}$, rather than just its Jordan structure. This seems very closely related to the subject of orientation theory, very nicely described in \cite{AS}. From the perspective of the topos approach, the natural question becomes whether orientations can be encoded by presheaves (contravariant, $\operatorname{Set}$-valued functors) over $AbSub~\mathcal{M}$, or maybe by covariant functors.

\section{Acknowledgements} 

We wish to thank Bruce Olberding for several helpful discussions related to abelian $C^*$-algebras; Daniel Marsden and Chris Heunen for comments; and the referee for carefully reading the paper and providing several suggestions.

\end{document}